\documentclass{article}

\usepackage{arxiv}
\usepackage[numbers,sort&compress]{natbib}
\usepackage{tabularray}
\usepackage{adjustbox}
 \usepackage{amsmath}
\usepackage[utf8]{inputenc} 
\usepackage[T1]{fontenc}    
\usepackage{hyperref}       
\usepackage{url}            
\usepackage{booktabs}       
\usepackage{amsfonts}       
\usepackage{nicefrac}       
\usepackage{microtype}      
\usepackage{lipsum}		
\usepackage{graphicx}
\usepackage{natbib}
\usepackage{doi}

\title{Eye Feel You: A DenseNet-driven User State Prediction Approach}
\renewcommand{\headeright}{}
\renewcommand{\undertitle}{}
\renewcommand{\shorttitle}{}

\date{} 					

\author{
Kamrul Hasan$^{1}$,
Oleg V. Komogortsev$^{1}$\\[2mm]
$^{1}$Texas State University, San Marcos, Texas, USA\\[2mm]
\texttt{\{kamrul.hasan, ok\}@txstate.edu}
}

\renewcommand{\shorttitle}{Eye Feel You: A DenseNet-driven User State Prediction Approach}

\hypersetup{
pdftitle={A template for the arxiv style},
pdfsubject={q-bio.NC, q-bio.QM},
pdfauthor={David S.~Hippocampus, Elias D.~Striatum},
pdfkeywords={First keyword, Second keyword, More},
}

\begin{document}
\maketitle

\begin{abstract}
Subjective self-reports, collected with eye-tracking data, reveal perceived states like fatigue, effort, and task difficulty. However, these reports are costly to collect and challenging to interpret consistently in longitudinal studies. In this work, we focus on determining whether objective gaze dynamics can reliably predict subjective reports across repeated recording rounds in the eye-tracking dataset. We formulate subjective-report prediction as a supervised regression problem and propose a DenseNet-based deep learning regressor that learns predictive representations from gaze velocity signals. We conduct two complementary experiments to clarify our aims. First, the cross-round generalization experiment tests whether models trained on earlier rounds transfer to later rounds, evaluating the models' ability to capture longitudinal changes. Second, cross-subject generalization tests models' robustness by predicting subjective outcomes for new individuals. These experiments aim to reduce reliance on hand-crafted feature designs and clarify which states of subjective experience systematically appear in oculomotor behavior over time.
\end{abstract}

\keywords{Eye Tracking, Subjective Report Prediction, Gaze Dynamics, Longitudinal Eye-movement Analysis, Cross-round Generalization, Cross-subject Generalization, Deep Regression, DenseNet, Oculomotor Features}

\section{Introduction}
Eye-tracking (ET) technology \cite{lai2013review} is becoming increasingly prominent for expanding the capabilities and applications of interactive and human-centered systems \cite{gardony2020eye}. ET applications include foveated rendering in AR/VR \cite{patney2016towards}, gaze-based interaction \cite{piumsomboon2017exploring}, user authentication \cite{lohr2022ekyt, lohr2025gaze}, and health and safety \cite{meissner2019promise, haller2022eye, kadhim2023implementation, liu2025state}. As ET shifts from controlled lab studies to repeated use and real-world deployments \cite{fong2016making}, researchers examine not only where users look but also how they feel while performing tasks, such as fatigue, mental effort, comfort, and task difficulty. These internal states can influence ET systems' performance in several ways, including altering system accuracy and reliability, affecting user interaction, changing perceived usability, and impacting the likelihood of long-term adoption \cite{di2016gaze, yang2025cross}.

To achieve optimal outcomes from ET-based applications, it is important to consider both objective gaze signals and corresponding subjective self-reports \cite{zhang2024gaze}. Subjective reports are broadly applicable, as they capture user experiences and perceived demands that are not directly observable from gaze behavior alone \cite{albert2010reliability}. However, subjective measures can be challenging to interpret consistently across repeated rounds. Additionally, context, memory effects, and differences in how people use rating scales can affect self-reports, adding time and burden to experiments. These challenges become more pronounced in longitudinal or multi-session studies \cite{qian2025we}.

Despite the importance of subjective reports and outcomes, many gaze datasets and modeling pipelines emphasize objective signals and task performance, providing limited support for learning or validating relationships between gaze dynamics and subjective experience \cite{thilderkvist2024current}. This gap motivates us to propose a computational approach that connects objective oculomotor behavior to subjective measures \cite{bilbao2021objective}. This approach enables scalable analyses and allows for understanding how perceived states change over time, especially when repeated self-report collection is impractical.

In this paper, we address the problem of predicting subjective states (e.g., fatigue, task difficulty, general comfort) from objective gaze-derived features to better understand how subjective reports manifest in eye movement dynamics across rounds. To begin, we formulate the subjective ratings estimation task as a supervised multi-target regression task \cite{nasteski2017overview}. Then, we propose a pre-activation DenseNet-based deep learning framework that extracts predictive features from gaze velocity signals \cite{huang2017densely}. Subsequently, we introduce a regressor head module that predicts the subjective scores. Overall, this approach reduces reliance on handcrafted feature designs and may increase modeling flexibility and accuracy. Lastly, we evaluate the approach across two experimental settings that assess longitudinal stability and robustness to individual differences for cross-round and cross-subject generalization. Through these evaluations, we clarify when gaze-based approaches offer reliable insights into subjective experience and when model personalization may be necessary. The specific contributions of our study are threefold:
\begin{enumerate}
\item We formulate subjective reports and eye-movement dynamics as a multi-target regression problem to understand subjective states across sessions.
\item We propose a pre-activation DenseNet-based deep regression framework and a regressor head module that directly learns from gaze-derived signals and predicts subjective scores.
\item We conduct two complementary experiments, including targeting cross-round generalization and targeting generalization to unseen individuals, to understand longitudinal effects from person-specific variability.
\end{enumerate}

\section{Related Work}
\subsection{Eye Tracking as an Objective Signal for Interactive Systems}
Eye tracking has progressed from an offline analysis tool to a practical sensing modality for interactive and adaptive systems \cite{duchowski2018gaze}. In the human-computer interaction (HCI) domain \cite{majaranta2014eye}, gaze-based applications cover use cases that range from usability assessment and expertise analysis to gaze-contingent displays (e.g., foveated rendering) and gaze-based selection \cite{plopski2022eye}. For example, \citet{albert2017latency} investigated how eye-tracking latency affects perceptual quality in VR foveated rendering. Later, \citet{matthews2020rendering} reviewed eye-tracking–based rendering optimizations for VR and survey techniques such as multiview rendering, learning-based foveated streaming, and variable- and multi-rate shading, and discussed opportunities enabled by consumer-grade eye tracking. Furthermore, some studies \cite{hu2025exploring, akshay2023ibehave, shevtsova2023machine, chen2021adaptive, schweigert2019eyepointing} have explored gaze as a rich behavioral stream that reflects underlying cognitive and perceptual processes, thus motivating learning-based approaches to model higher-level subjective states from eye-movement dynamics.

\subsection{Subjective State Measurement and Gaze-Based Prediction}
Subjective self-reports \cite{zentner2010self, ferreira2019subjective, faber2018automated} remain a primary method for quantifying internal states such as workload, effort, and fatigue. NASA-TLX \cite{hart1988development} is one of the most widely adopted instruments for subjective workload assessment and has been extensively used across lab and applied evaluations. However, repeatedly collecting self-reports can be burdensome and introduce variability due to context and individual differences in scale use (i.e., subjective ratings), motivating methods that infer subjective outcomes from objective behavior.

In addition, recent gaze datasets have provided paired eye-tracking signals and subjective annotations, enabling supervised prediction of cognitive state. For example, the COLET \cite{ktistakis2022colet} dataset explicitly labels cognitive workload using NASA-TLX and supports benchmarking of workload estimation from gaze features. Similarly, the GazeBase \cite{griffith2021gazebase} dataset provides a subjective fatigue rating on a scale of 1-7 for six different tasks, where 1 indicates no fatigue, and 7 indicates maximum fatigue. Previous studies \cite{pandey2021temporal, zemblys2019gazenet, cheng2024appearance} have investigated both feature-based machine learning and deep learning approaches to extract the temporal structure of eye movements and understand the gaze patterns. However, emphasizing generalization across tasks, rounds, and individuals remains a significant challenge for reliable subjective-state inference.

\subsection{Learning-Based Regression for Subjective-State Prediction}
Machine learning-based approaches \cite{matloff2017statistical, alnuaimi2024overview, kadam2020regression} are ideal for classification and regression analysis \cite{hoffmann2019benchmarking}, as they can learn continuous mappings from noisy behavioral signals to subjective outcomes while modeling nonlinear relationships. In eye-tracking research, learning-based approaches \cite{kaczorowska2021interpretable, skaramagkas2021cognitive, li2025applications} have been extensively used to infer cognitive and affective states (e.g., workload or fatigue) from gaze- and pupil-derived metrics \cite{skaramagkas2021review, susmitha2024mitigating, basnet2025real}. For example, \citet{kashevnik2024intelligent} proposed a real-time eye-tracking–based method for detecting mental fatigue, identifying key gaze features using a random forest classifier. Later, \citet{bitkina2021ability} examined that eye-tracking metrics can be used to predict perceived driving workload, supporting the feasibility of gaze-to-subjective modeling in realistic tasks. However, many classical pipelines also rely on handcrafted features that may discard the informative temporal structure present in continuous gaze dynamics \cite{rudmann2003eyetracking, saiz2021analysis, ye2023supporting}. To the best of our knowledge, predicting subjective state scores from raw gaze dynamics remains underexplored in the current literature. To address this research gap, we adopt a DenseNet-based \cite{huang2017densely} multi-target regressor that leverages dense connectivity to promote feature reuse and stable gradient flow, well-suited for learning robust representations from complex eye-movement signals.

\section{Proposed Architecture}
\begin{figure*}[ht]
    \centering
\includegraphics[width=\linewidth]{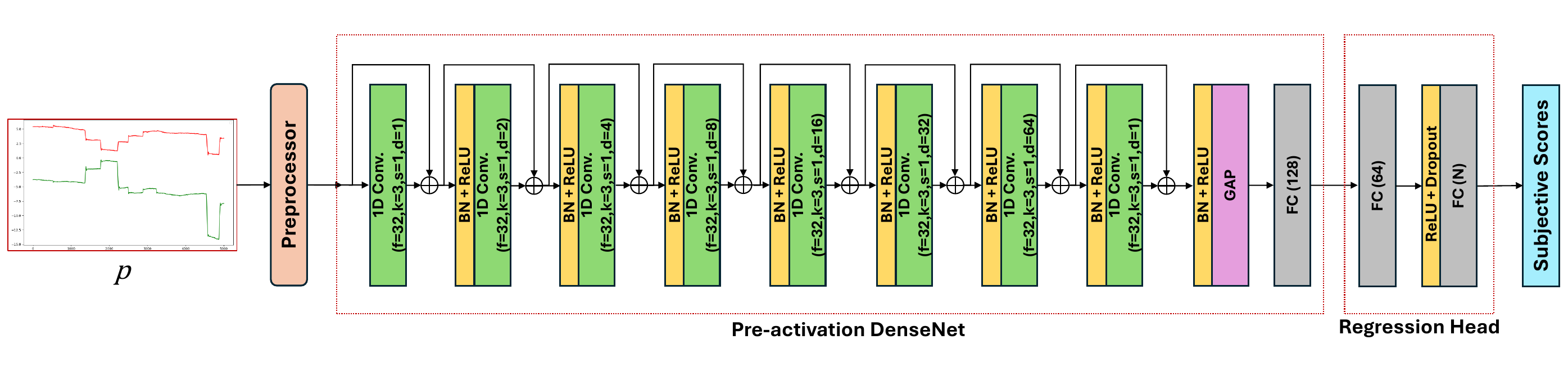}
    \caption{Overview of the architecture, where the positional signal $p$ is processed by the Preprocessor, yielding a velocity signal, which is then fed to the Pre-activation DenseNet. Here, each convolutional layer has a kernel size $k = 3$, a stride $s = 1$, and a varying dilation rate $d$. Later, Regression Head takes the output from DenseNet and predicts Subject Scores, where $N$ denotes the number of predicted scores.}
    \label{fig:main_architecture}
\end{figure*}

In this section, we present the overall architecture used for the multi-target regression task, which leverages objective gaze velocity signals to predict subjective scores. As shown in Figure \ref{fig:main_architecture}, the pipeline begins with a preprocessor module that transforms raw gaze positions into a normalized velocity signal. The resulting signal is then fed into a pre-activation DenseNet-based backbone for representation learning, followed by a regression head that predicts subjective scores.

\subsection{Data Preprocessing}
Let the gaze sequence be $p \in \mathbb{R}^{S \times d}$, where $S$ is the number of samples and $d=2$ for horizontal and vertical gaze positions, with samples measured in milliseconds $(ms)$. In the preprocessor module, each gaze sequence is cropped to the first 50 seconds and downsampled from 1000 Hz to 100 Hz, yielding $S=5000$ samples. The positional signal ($^{\circ}$) is then converted to velocity ($^{\circ}/s$) using a Savitzky–Golay \cite{savitzky1964smoothing} differentiation filter (SGDF) with polynomial order 2, first derivative, and window size 7:

\begin{equation}
    \begin{split}
v &= \mathit{SGDF}(p)
    \end{split}
\end{equation}

\noindent where $v$ denotes the resulting velocity signal. Later, any invalid velocity samples (e.g., NaNs) are set to zero. Finally, to reduce the impact of noise and spikes, velocities are clipped to [-1000,  1000] °/s, then linearly rescaled to [-90, 90] and normalized to [-1, 1] using a sine-based transformation that constrains amplitude while preserving relative dynamics. The resulting normalized velocity signal is then used as input to the DenseNet-based regressor for feature extraction, which is subsequently fed to the regressor head for subjective score prediction. Additionally, to use the subjective reports as regression targets, we aligned each preprocessed gaze recording with its corresponding survey entry using the subject identifier and the experimental condition (session/task). This yields paired samples $(x,y)$, where $x$ is the normalized velocity sequence and $y$ is the vector of subjective ratings used to supervise the regressor.

\subsection{DenseNet Feature Extractor and Regression Head}
We propose a pre-activation DenseNet-based regression framework as presented in Figure \ref{fig:main_architecture}. It maps the processed gaze velocity signal to continuous subjective-report targets. It follows the DenseNet principle of dense connectivity, where the concatenation of feature maps from all preceding layers within a dense block serves as each layer's input. This arrangement facilitates feature reuse and improves gradient flow.

As shown in Figure \ref{fig:main_architecture}, the global average pooling (GAP) layer and all convolution layers except the first are preceded by batch normalization (BN) and Rectified Linear Unit (ReLU) activation. This ordering results in the sequence BN-ReLU-Convolution. Since BN is used, convolution layers do not include an additive bias term. Additionally, we employed a growth rate of 32, with each convolutional layer outputting 32 feature maps that are concatenated with the previous feature maps. Here, convolution layers (i.e., $n$) numbered from 1 to 8 were used, with kernel size $k = 3$ and stride $s = 1$. To efficiently increase temporal context, we applied exponentially increasing dilations (i.e., $d_n = 2^{n-1} \pmod{7}$) and employed necessary zero-padding (i.e., $ p_n = d_n$) on both sides to preserve sequence length. Finally, the receptive field of layer $n$, denoted as $r_n$, can be present as follows:

\begin{equation}
r_n = 1 + \sum_{i=1}^{n} d_i(k_i - 1)
\end{equation}

\noindent With the above settings, the final convolution layer achieves a maximum receptive field of $r_8 = 257$ time steps from the input.

After extracting features from the convolutional layers, a GAP layer then aggregates temporal features into a compact embedding. This representation is subsequently passed to a fully connected (FC) layer, which serves as the input to the regression head module. Finally, the regression head consists of an FC layer followed by a ReLU-Dropout-FC layer and predicts multi-target regression scores (i.e., $N$) based on the experimental setting.

\section{Experiment}
\subsection{Hardware \& Software}
All necessary experiments were conducted on a Linux 24.04 operating system (OS) with Python 3.9.12, CUDA 12.4, PyTorch 2.1.0, PyTorch-Lightning 1.9.5, Numpy, and other necessary libraries on a single NVIDIA RTX A6000 GPU (48 GB GDDR6). And to ensure reproducibility, we will release the complete source code.

\subsection{Dataset}
For our experiments, we use the publicly available GazeBase \cite{griffith2021gazebase} dataset, which contains objective gaze recordings. GazeBase was collected using an EyeLink 1000 eye tracker and has monocular (left-eye) gaze position signals sampled at 1000 Hz. The dataset spans nine recording rounds over approximately 37 months, includes 322 participants, and contains 12,334 eye-tracking recordings. In each round, participants completed a battery of seven tasks: fixation (FXS), horizontal saccades (HSS), random oblique saccades (RAN), reading (TEX), free viewing of cinematic video (VD1 and VD2), and a gaze-driven game (BLG), and each task has two sessions.

Our study utilizes gaze signals and the subjective ratings dataset from GazeBase, collected alongside the recordings. We employed two experimental settings: cross-round generalization (i.e., score prediction for a known subject) and cross-subject generalization (i.e., score prediction for an unknown subject). In the known-subject setting, we utilize per-session/per-task ratings from rounds 2–9 for mental tiredness (Mentally), eye tiredness (TiredEyes), and overall difficulty (OverDiff), each rated on a 1-7 scale. In the unknown-subject setting, we utilize survey scores with six outcomes: general\_comfort, shoulder\_fatigue, neck\_fatigue, eye\_fatigue, physical\_effort, and mental\_effort. It is also rated on a 1–7 scale and reported in both the between-session and after-session questionnaires. Our work predicts (i) three subjective scores in cross-round generalization and (ii) six survey-based scores in cross-subject generalization in two separate experiments.

\subsection{Training}
We train the DenseNet-based framework with a robust regression loss (Smooth L1) and optimize using Adam with a batch size of 16 and a learning rate of 3e-4. We also employed regularization strategies, including dropout, weight decay, and early stopping using a validation split. Overall, we trained the model for 50 epochs. In the cross-subject generalization experiment (i.e., for known subjects), the model predicts three subjective ratings: overall difficulty, mental tiredness, and eye tiredness. Here, 80\% of the data from round 2 is used for training, 20\% for validation, and rounds 3 and 4 for testing. In the experiment for cross-round generalization (i.e., for unknown subjects), the model predicts six scores, including general\_comfort, shoulder\_fatigue, neck\_fatigue, eye\_fatigue, physical\_effort, and mental\_effort, where subjects from ($ \text{round 1} \cup \text{round 2} $) are used for training (80\%) and validation (20\%), and subjects from ($ \text{round 1} \cap \text{round 2} $) are used for testing. To support fair comparisons, we also report a baseline, including a global mean prediction, computed by taking the per-dimension mean of all training labels and using this vector as the prediction for every sample.

\subsection{Testing}
We evaluate prediction performance using standard regression and agreement metrics. For evaluation, we report Mean Absolute Error (MAE) and Root Mean Squared Error (RMSE) to summarize average deviation and penalize larger errors, respectively. We also report Pearson correlation (r) to measure linear association between predictions and ground truth, and the coefficient of determination (R²) to quantify explained variance relative to a mean predictor. In addition, as subjective ratings lie on a discrete 1–7 scale, we include an agreement metric based on rounding predictions to the nearest integer (i.e., exact accuracy), defined as the fraction of samples where the rounded prediction exactly matches the true rating. All metrics are reported separately for the known-subject and unknown-subject settings to distinguish cross-round generalization from cross-subject generalization.

\section{Results}
In this section, we report experimental performances using MAE, RMSE, Pearson correlation (r), R², and exact accuracy, where lower MAE/RMSE indicates smaller prediction error and higher r/R²/accuracy indicates variance and stronger agreement. Across both experiments, we compare the DenseNet-based framework against a simple global mean baseline to quantify the value added by learned gaze representations.

\subsection{Performance on Cross-rounds}
As presented in Tables \ref{tab:experiment1_overall} and \ref{tab:experiment1_specific}, in the known-subject setting (i.e., cross-round evaluation), DenseNet consistently reduces MAE and substantially improves exact accuracy compared to the global mean baseline. In round 3, compared to the global mean, DenseNet reduces the MAE from 0.78 to 0.64 and increases the accuracy from 0.22 to 0.60. In round 4, MAE decreases from 0.79 to 0.61, while accuracy increases from 0.18 to 0.65.

A breakdown by target in Table \ref{tab:experiment1_specific} shows that DenseNet improves MAE for all three ratings (overall difficulty, mental tiredness, and eye tiredness) in both rounds. DenseNet also yields higher correlations (e.g., r = 0.19–0.35), indicating that it captures meaningful ordering trends in the subjective labels. However, R² remains negative across targets, and RMSE is sometimes slightly higher than the global mean, which suggests that while the model improves average error and discrete agreement, it does not consistently explain variance better than a mean predictor and may still produce occasional larger deviations.

\begin{table}[h]
\centering
\caption{Overall performance on Experiment 1 (i.e., known-subjects) for round 3 and round 4, comparing Global Mean and DenseNet using MAE, RMSE, and exact Accuracy. Here, the symbols '↓' and '↑' indicate that lower and higher values are better, respectively.}
\label{tab:experiment1_overall}
\begin{adjustbox}{width=0.8\textwidth}
{\small
\renewcommand{\arraystretch}{1}
\begin{tblr}{
  cells = {c},
  cell{1}{1} = {r=2}{},
  cell{1}{2} = {c=3}{},
  cell{1}{5} = {c=3}{},
  hline{1,3,5} = {-}{},
  hline{2} = {2-7}{},
}
\textbf{Method}      & \textbf{Round 3} &                 &                     & \textbf{Round 4} &                 &                     \\
                     & \textbf{~MAE ↓}  & \textbf{~RMSE ↓} & \textbf{~Accuracy ↑} & \textbf{~MAE ↓}  & \textbf{~RMSE ↓} & \textbf{~Accuracy↑} \\
\textbf{Global Mean} & 0.78             & 1.00            & 0.22                & 0.79             & 1.02            & 0.18                \\
\textbf{DenseNet}    & 0.64             & 1.04            & 0.60                & 0.61             & 1.03            & 0.65                
\end{tblr}
}
\end{adjustbox}
\end{table}

\begin{table}[h]
\centering
\caption{Per-target evaluation for Experiment 1 (i.e., known-subject) on round 3 and round 4 for Overall Difficulty, Mental Tiredness, and Eye Tiredness, reported with MAE, RMSE, Pearson r, and R² for Global Mean and DenseNet. Here '-' denotes undefined Pearson (r) for Global Mean because the prediction is constant (zero variance) and the symbols '↓' and '↑' indicate that lower and higher values are better, respectively.}
\label{tab:experiment1_specific}
\begin{adjustbox}{width=0.9\textwidth}
{\small
\renewcommand{\arraystretch}{1}
\begin{tblr}{
  cells = {c},
  cell{1}{1} = {r=2}{},
  cell{1}{2} = {r=2}{},
  cell{1}{3} = {c=4}{},
  cell{1}{7} = {c=4}{},
  cell{3}{1} = {r=2}{},
  cell{5}{1} = {r=2}{},
  cell{7}{1} = {r=2}{},
  hline{1,3,5,7,9} = {-}{},
  hline{2} = {3-10}{},
}
\textbf{Metric}    & \textbf{Method}      & \textbf{Round 3} &                &             &              & \textbf{Round 4} &                &             &              \\
                   &                      & \textbf{MAE ↓}    & \textbf{RMSE ↓} & \textbf{r ↑} & \textbf{R² ↑} & \textbf{MAE ↓}    & \textbf{RMSE ↓} & \textbf{r ↑} & \textbf{R² ↑} \\
\textbf{OverDiff}  & \textbf{Global Mean} & 0.71             & 0.90           & -        & 0.00         & 0.71             & 0.91           & 0.00        & -0.01        \\
                   & \textbf{DenseNet}    & 0.56             & 0.94           & 0.24        & -0.10        & 0.53             & 0.93           & 0.22        & -0.07        \\
\textbf{Mentally}  & \textbf{Global} Mean & 0.71             & 0.89           & -        & 0.00         & 0.72             & 0.90           & -        & -0.02        \\
                   & \textbf{DenseNet}    & 0.56             & 0.94           & 0.20        & -0.11        & 0.52             & 0.93           & 0.19        & -0.06        \\
\textbf{TiredEyes} & \textbf{Global} Mean & 0.93             & 1.19           & 0.00        & -0.01        & 0.95             & 1.21           & 0.00        & -0.02        \\
                   & \textbf{DenseNet}    & 0.81             & 1.21           & 0.35        & -0.04        & 0.78             & 1.20           & 0.31        & -0.02        
\end{tblr}
}
\end{adjustbox}
\end{table}

\subsection{Performance on Cross-subjects}
As shown in Tables \ref{tab:experiment2_overall} and \ref{tab:experiment2_specific}, DenseNet yields only slight improvements over the global mean baseline in the unknown-subject (i.e., cross-subject) setting. Between sessions, MAE shifts from 1.36 to 1.35, and accuracy from 0.21 to 0.22. In after sessions, MAE remains 1.35, while accuracy increases from 0.21 to 0.22, as reported in Table \ref{tab:experiment2_overall}.

Similarly, each task-wise breakdown in Table \ref{tab:experiment2_specific} shows that DenseNet provides small, target-dependent changes. For some outcomes, such as general comfort and neck fatigue in the between-session phase, MAE improves slightly. For others, performance is nearly identical to the baseline. Correlations remain close to zero, and R² values are small, reflecting limited cross-person transfer from gaze dynamics to subjective scores.

\begin{table}[h]
\centering
\caption{Overall performance on Experiment 1 (i.e., unknown-subjects) for Between Sessions and After Sessions, comparing Global Mean and DenseNet using MAE, RMSE, and exact Accuracy. Here, the symbols '↓' and '↑' indicate that lower and higher values are better, respectively.}
\label{tab:experiment2_overall}
\begin{adjustbox}{width=0.8\textwidth}
{\small
\renewcommand{\arraystretch}{1}
\begin{tblr}{
  cells = {c},
  cell{1}{1} = {r=2}{},
  cell{1}{2} = {c=3}{},
  cell{1}{5} = {c=3}{},
  hline{1,5} = {-}{},
  hline{2-3} = {2-7}{},
}
\textbf{Method}      & \textbf{Between Sessions} &                 &                     & \textbf{After Sessions} &                 &                     \\
                     & \textbf{~MAE ↓}           & \textbf{~RMSE ↓} & \textbf{~Accuracy ↑} & \textbf{~MAE ↓}         & \textbf{~RMSE ↓} & \textbf{~Accuracy ↑} \\
\textbf{Global Mean} & 1.36                      & 1.67            & 0.21                & 1.35                    & 1.67            & 0.21                \\
\textbf{DenseNet}    & 1.35                      & 1.69            & 0.22                & 1.35                    & 1.68            & 0.22                
\end{tblr}
}
\end{adjustbox}
\end{table}

\begin{table}[h]
\centering
\caption{Per-target evaluation for Experiment 2 (i.e., unknown-subject) on Between Sessions and After Sessions for general\_comfort, shoulder\_fatigue, neck\_fatigue, eyes\_fatigue, physical\_effort, and mental\_effort, reported with MAE, RMSE, Pearson r, and R² for Global Mean and DenseNet. Here '-' denotes undefined Pearson (r) for Global Mean because the prediction is constant (zero variance) and the symbols '↓' and '↑' indicate that lower and higher values are better, respectively.}
\label{tab:experiment2_specific}
\begin{adjustbox}{width=0.9\textwidth}
{\small
\renewcommand{\arraystretch}{1}
\begin{tblr}{
  cells = {c},
  cell{1}{1} = {r=2}{},
  cell{1}{2} = {r=2}{},
  cell{1}{3} = {c=4}{},
  cell{1}{7} = {c=4}{},
  cell{3}{1} = {r=2}{},
  cell{5}{1} = {r=2}{},
  cell{7}{1} = {r=2}{},
  cell{9}{1} = {r=2}{},
  cell{11}{1} = {r=2}{},
  cell{13}{1} = {r=2}{},
  hline{1,3,5,7,9,11,13,15} = {-}{},
  hline{2} = {3-10}{},
}
\textbf{Metric}            & \textbf{Method}      & \textbf{Between Sessions} &                &             &              & \textbf{After Sessions} &                &             &              \\
                           &                      & \textbf{MAE ↓}             & \textbf{RMSE ↓} & \textbf{r ↑} & \textbf{R² ↑} & \textbf{MAE ↓}           & \textbf{RMSE ↓} & \textbf{r ↑} & \textbf{R² ↑} \\
\textbf{general\_comfort}  & \textbf{Global Mean} & 1.01                      & 1.26           & 0.00        & -0.06        & 1.02                    & 1.31           & -        & -0.01        \\
                           & \textbf{DenseNet}    & 1.00                      & 1.25           & 0.04        & -0.03        & 1.02                    & 1.31           & -0.02       & -0.01        \\
\textbf{shoulder\_fatigue} & \textbf{Global Mean} & 1.41                      & 1.67           & 0.00        & 0.00         & 1.33                    & 1.61           & 0.00        & -0.02        \\
                           & \textbf{DenseNet}    & 1.39                      & 1.67           & 0.06        & -0.01        & 1.33                    & 1.60           & 0.04        & 0.00         \\
\textbf{neck\_fatigue}     & \textbf{Global Mean} & 1.40                      & 1.65           & 0.00        & -0.01        & 1.38                    & 1.64           & 0.00        & -0.04        \\
                           & \textbf{DenseNet}    & 1.38                      & 1.64           & 0.01        & 0.00         & 1.36                    & 1.63           & 0.04        & -0.02        \\
\textbf{eyes\_fatigue}     & \textbf{Global Mean} & 1.25                      & 1.50           & 0.00        & -0.01        & 1.28                    & 1.55           & 0.00        & -0.02        \\
                           & \textbf{DenseNet}    & 1.25                      & 1.50           & 0.00        & -0.02        & 1.28                    & 1.55           & 0.05        & -0.02        \\
\textbf{physical\_effort}  & \textbf{Global Mean} & 1.53                      & 1.96           & -        & -0.07        & 1.52                    & 1.93           & 0.00        & -0.06        \\
                           & \textbf{DenseNet}    & 1.50                      & 2.03           & 0.07        & -0.16        & 1.52                    & 1.97           & 0.06        & -0.10        \\
\textbf{mental\_effort}    & \textbf{Global Mean} & 1.57                      & 1.91           & -        & -0.09        & 1.55                    & 1.88           & 0.00        & -0.09        \\
                           & \textbf{DenseNet}    & 1.58                      & 1.92           & 0.10        & -0.11        & 1.58                    & 1.94           & -0.03       & -0.17        
\end{tblr}
}
\end{adjustbox}
\end{table}

\section{Discussion}
The two different experiments highlight a clear difference between within-person longitudinal generalization (i.e., how well the model performs on the same person over time) and cross-person generalization (i.e., how well the model performs across different people). In the known-subject setting, our proposed DenseNet improves MAE (i.e., mean absolute error) and discrete agreement (i.e., accuracy), suggesting that gaze dynamics contain subject-consistent features related to perceived tiredness and difficulty that transfer across rounds. More specifically, for overall difficulty and mental tiredness, MAE ranges from 0.52 to 0.56 across rounds 3 and 4. The moderate increases in correlation further support the deep-learning model's ability to capture meaningful trends in the labels. At the same time, negative R² (i.e., coefficient of determination) and occasional increases in RMSE (i.e., root mean squared error) also imply that subjective ratings remain noisy and difficult to explain in a variance-sensitive manner, likely due to limited label variance across tasks, individual scale-use variability, and the coarse 1–7 scoring resolution.

In the unknown-subject setting from Tables \ref{tab:experiment2_overall} and \ref{tab:experiment2_specific}, the weak gains suggest that known individual differences dominate the gaze–subjective relationship. This is expected because eye-movement patterns and how participants use subjective scales can vary substantially across people. These results motivate directions such as subject-aware normalization, personalization/calibration, domain adaptation, or hybrid models that combine temporal gaze dynamics with additional context (task identity, session factors) to improve effectiveness for unseen individuals.

\section{Conclusion}
In this work, we investigated predicting subjective reports from objective eye-movement signals using a DenseNet-based regression framework for two evaluation settings. For known subjects across rounds, the model shows clear benefits, improving average errors and discrete agreement for mental fatigue, eye fatigue, and overall difficulty. In contrast, unknown-subject prediction remains challenging, with performance close to simple mean baselines and weak correlations indicating limited cross-person generalization. These findings indicate that gaze dynamics may facilitate longitudinal modeling of subjective experience within individuals; however, effective out-of-subject inference probably necessitates personalization or supplementary modeling strategies.

\section*{Privacy and Ethics Statement}
This study uses anonymized, publicly available eye-movement data to train and test the model, posing minimal societal risk. We encourage responsible application to avoid potential misuse. Our work supports ethical, privacy-conscious advancements in eye-tracking research.

\bibliographystyle{unsrtnat}
\bibliography{references}  

@String{Computing = "Computing" }

@String{Computer = "{IEEE} Computer" }

@String{Springer = "Springer-Verlag" }

@article{griffith2021gazebase,
  title={GazeBase, a large-scale, multi-stimulus, longitudinal eye movement dataset},
  author={Griffith, Henry and Lohr, Dillon and Abdulin, Evgeny and Komogortsev, Oleg},
  journal={Scientific Data},
  volume={8},
  number={1},
  pages={184},
  year={2021},
  publisher={Nature Publishing Group UK London}
}

@article{meissner2019promise,
  title={The promise of eye-tracking methodology in organizational research: A taxonomy, review, and future avenues},
  author={Mei{\ss}ner, Martin and Oll, Josua},
  journal={Organizational Research Methods},
  volume={22},
  number={2},
  pages={590--617},
  year={2019},
  publisher={Sage Publications Sage CA: Los Angeles, CA}
}

@article{haller2022eye,
  title={Eye-tracking based classification of Mandarin Chinese readers with and without dyslexia using neural sequence models},
  author={Haller, Patrick and S{\"a}uberli, Andreas and Kiener, Sarah Elisabeth and Pan, Jinger and Yan, Ming and J{\"a}ger, Lena},
  journal={arXiv preprint arXiv:2210.09819},
  year={2022}
}

@article{lohr2022ekyt,
  title={Eye know you too: Toward viable end-to-end eye movement biometrics for user authentication},
  author={Lohr, Dillon and Komogortsev, Oleg V},
  journal={IEEE Transactions on Information Forensics and Security},
  volume={17},
  pages={3151--3164},
  year={2022},
  publisher={IEEE}
}

@article{patney2016towards,
  title={Towards foveated rendering for gaze-tracked virtual reality},
  author={Patney, Anjul and Salvi, Marco and Kim, Joohwan and Kaplanyan, Anton and Wyman, Chris and Benty, Nir and Luebke, David and Lefohn, Aaron},
  journal={ACM Transactions On Graphics (TOG)},
  volume={35},
  number={6},
  pages={1--12},
  year={2016},
  publisher={ACM New York, NY, USA}
}

@article{lohr2025gaze,
  title={Gaze Authentication: Factors Influencing Authentication Performance},
  author={Lohr, Dillon and Proulx, Michael J and Raju, Mehedi Hasan and Komogortsev, Oleg V},
  journal={arXiv preprint arXiv:2509.10969},
  year={2025}
}

@article{savitzky1964smoothing,
  title={Smoothing and differentiation of data by simplified least squares procedures.},
  author={Savitzky, Abraham and Golay, Marcel JE},
  journal={Analytical chemistry},
  volume={36},
  number={8},
  pages={1627--1639},
  year={1964},
  publisher={ACS Publications}
}

@article{qian2025we,
  title={Why do we need high-fidelity synthetic eye movement data and how should they look like?},
  author={Qian, C Stella and Aziz, Samantha and Hasan, Kamrul and Komogortsev, Oleg V},
  journal={bioRxiv},
  pages={2025--12},
  year={2025},
  publisher={Cold Spring Harbor Laboratory}
}

@article{duchowski2018gaze,
  title={Gaze-based interaction: A 30 year retrospective},
  author={Duchowski, Andrew T},
  journal={Computers \& Graphics},
  volume={73},
  pages={59--69},
  year={2018},
  publisher={Elsevier}
}

@incollection{majaranta2014eye,
  title={Eye tracking and eye-based human--computer interaction},
  author={Majaranta, P{\"a}ivi and Bulling, Andreas},
  booktitle={Advances in physiological computing},
  pages={39--65},
  year={2014},
  publisher={Springer}
}

@article{plopski2022eye,
  title={The eye in extended reality: A survey on gaze interaction and eye tracking in head-worn extended reality},
  author={Plopski, Alexander and Hirzle, Teresa and Norouzi, Nahal and Qian, Long and Bruder, Gerd and Langlotz, Tobias},
  journal={ACM Computing Surveys (CSUR)},
  volume={55},
  number={3},
  pages={1--39},
  year={2022},
  publisher={ACM New York, NY}
}

@article{albert2017latency,
  title={Latency requirements for foveated rendering in virtual reality},
  author={Albert, Rachel and Patney, Anjul and Luebke, David and Kim, Joohwan},
  journal={ACM Transactions on Applied Perception (TAP)},
  volume={14},
  number={4},
  pages={1--13},
  year={2017},
  publisher={ACM New York, NY, USA}
}

@inproceedings{matthews2020rendering,
  title={Rendering optimizations for virtual reality using eye-tracking},
  author={Matthews, Sage L and Uribe-Quevedo, Alvaro and Theodorou, Alexander},
  booktitle={2020 22nd symposium on virtual and augmented reality (SVR)},
  pages={398--405},
  year={2020},
  organization={IEEE}
}

@article{hu2025exploring,
  title={Exploring and modeling the effects of eye-tracking accuracy and precision on gaze-based steering in virtual environments},
  author={Hu, Xuning and Zhang, Yichuan and Wei, Yushi and Zhang, Liangyuting and Li, Yue and Stuerzlinger, Wolfgang and Liang, Hai-Ning},
  journal={IEEE Transactions on Visualization and Computer Graphics},
  year={2025},
  publisher={IEEE}
}

@inproceedings{akshay2023ibehave,
  title={ibehave: Behaviour analysis using eye gaze metrices},
  author={Akshay, S and Kavya Bijith, P and Sanjana, S and Amudha, Joseph},
  booktitle={International Conference on Pattern Recognition and Machine Intelligence},
  pages={260--269},
  year={2023},
  organization={Springer}
}

@inproceedings{shevtsova2023machine,
  title={Machine learning for gaze-based selection: Performance assessment without explicit labeling},
  author={Shevtsova, Yulia G and Vasilyev, Anatoly N and Shishkin, Sergei L},
  booktitle={International Conference on Human-Computer Interaction},
  pages={311--322},
  year={2023},
  organization={Springer}
}

@inproceedings{chen2021adaptive,
  title={An adaptive model of gaze-based selection},
  author={Chen, Xiuli and Acharya, Aditya and Oulasvirta, Antti and Howes, Andrew},
  booktitle={Proceedings of the 2021 CHI Conference on Human Factors in Computing Systems},
  pages={1--11},
  year={2021}
}

@incollection{schweigert2019eyepointing,
  title={Eyepointing: A gaze-based selection technique},
  author={Schweigert, Robin and Schwind, Valentin and Mayer, Sven},
  booktitle={Proceedings of mensch und computer 2019},
  pages={719--723},
  year={2019}
}

@incollection{hart1988development,
  title={Development of NASA-TLX (Task Load Index): Results of empirical and theoretical research},
  author={Hart, Sandra G and Staveland, Lowell E},
  booktitle={Advances in psychology},
  volume={52},
  pages={139--183},
  year={1988},
  publisher={Elsevier}
}

@incollection{ferreira2019subjective,
  title={Subjective and objective measures},
  author={Ferreira, Hugo Alexandre and Saraiva, Magda},
  booktitle={Emotional design in human-robot interaction: Theory, methods and applications},
  pages={143--159},
  year={2019},
  publisher={Springer}
}

@article{faber2018automated,
  title={An automated behavioral measure of mind wandering during computerized reading},
  author={Faber, Myrthe and Bixler, Robert and D’Mello, Sidney K},
  journal={Behavior Research Methods},
  volume={50},
  number={1},
  pages={134--150},
  year={2018},
  publisher={Springer}
}

@article{zentner2010self,
  title={Self-report measures and models},
  author={Zentner, Marcel and Eerola, Tuomas},
  journal={Handbook of music and emotion: Theory, research, applications},
  pages={187--221},
  year={2010}
}

@article{ktistakis2022colet,
  title={COLET: A dataset for COgnitive workLoad estimation based on eye-tracking},
  author={Ktistakis, Emmanouil and Skaramagkas, Vasileios and Manousos, Dimitris and Tachos, Nikolaos S and Tripoliti, Evanthia and Fotiadis, Dimitrios I and Tsiknakis, Manolis},
  journal={Computer Methods and Programs in Biomedicine},
  volume={224},
  pages={106989},
  year={2022},
  publisher={Elsevier}
}

@article{bitkina2021ability,
  title={The ability of eye-tracking metrics to classify and predict the perceived driving workload},
  author={Bitkina, Olga Vl and Park, Jaehyun and Kim, Hyun K},
  journal={International Journal of Industrial Ergonomics},
  volume={86},
  pages={103193},
  year={2021},
  publisher={Elsevier}
}

@article{kashevnik2024intelligent,
  title={Intelligent Human Operator Mental Fatigue Assessment Method Based on Gaze Movement Monitoring},
  author={Kashevnik, Alexey and Kovalenko, Svetlana and Mamonov, Anton and Hamoud, Batol and Bulygin, Aleksandr and Kuznetsov, Vladislav and Shoshina, Irina and Brak, Ivan and Kiselev, Gleb},
  journal={Sensors},
  volume={24},
  number={21},
  pages={6805},
  year={2024},
  publisher={MDPI}
}

@inproceedings{huang2017densely,
  title={Densely connected convolutional networks},
  author={Huang, Gao and Liu, Zhuang and Van Der Maaten, Laurens and Weinberger, Kilian Q},
  booktitle={Proceedings of the IEEE conference on computer vision and pattern recognition},
  pages={4700--4708},
  year={2017}
}

@article{lai2013review,
  title={A review of using eye-tracking technology in exploring learning from 2000 to 2012},
  author={Lai, Meng-Lung and Tsai, Meng-Jung and Yang, Fang-Ying and Hsu, Chung-Yuan and Liu, Tzu-Chien and Lee, Silvia Wen-Yu and Lee, Min-Hsien and Chiou, Guo-Li and Liang, Jyh-Chong and Tsai, Chin-Chung},
  journal={Educational research review},
  volume={10},
  pages={90--115},
  year={2013},
  publisher={Elsevier}
}

@inproceedings{gardony2020eye,
  title={Eye-tracking for human-centered mixed reality: promises and challenges},
  author={Gardony, Aaron L and Lindeman, Robert W and Bruny{\'e}, Tad T},
  booktitle={Optical architectures for displays and sensing in augmented, virtual, and mixed reality (AR, VR, MR)},
  volume={11310},
  pages={230--247},
  year={2020},
  organization={SPIE}
}

@inproceedings{piumsomboon2017exploring,
  title={Exploring natural eye-gaze-based interaction for immersive virtual reality},
  author={Piumsomboon, Thammathip and Lee, Gun and Lindeman, Robert W and Billinghurst, Mark},
  booktitle={2017 IEEE symposium on 3D user interfaces (3DUI)},
  pages={36--39},
  year={2017},
  organization={IEEE}
}

@article{liu2025state,
  title={State of eye-tracking technology research to enhance construction safety},
  author={Liu, Peng and Zhang, Chengyi and Arditi, David and Li, Huimin and Liang, Mengxuan},
  journal={Journal of Safety Research},
  volume={94},
  pages={317--328},
  year={2025},
  publisher={Elsevier}
}

@article{kadhim2023implementation,
  title={Implementation of eye-tracking technology to monitor clinician fatigue in routine clinical care: a feasibility study},
  author={Kadhim, Bashar and Khairat, Saif and Li, Fangyong and Gross, Isabel T and Nath, Bidisha and Hauser, Ronald G and Melnick, Edward R},
  journal={ACI Open},
  volume={7},
  number={01},
  pages={e1--e7},
  year={2023},
  publisher={Georg Thieme Verlag KG}
}

@article{di2016gaze,
  title={Gaze entropy reflects surgical task load},
  author={Di Stasi, Leandro L and Diaz-Piedra, Carolina and Rieiro, H{\'e}ctor and Sanchez Carrion, Jose M and Martin Berrido, Mercedes and Olivares, Gonzalo and Catena, Andr{\'e}s},
  journal={Surgical endoscopy},
  volume={30},
  number={11},
  pages={5034--5043},
  year={2016},
  publisher={Springer}
}

@article{yang2025cross,
  title={Cross-task cognitive workload estimation using eye tracking},
  author={Yang, Lin and Wang, Lei and Xu, Wenchang and Wang, Biao and Ren, Hanbin and Yang, Aijuan},
  journal={Signal, Image and Video Processing},
  volume={19},
  number={5},
  pages={354},
  year={2025},
  publisher={Springer}
}

@inproceedings{fong2016making,
  title={Making sense of mobile eye-tracking data in the real-world: A human-in-the-loop analysis approach},
  author={Fong, Allan and Hoffman, Daniel and Ratwani, Raj M},
  booktitle={Proceedings of the human factors and ergonomics society annual meeting},
  volume={60},
  number={1},
  pages={1569--1573},
  year={2016},
  organization={SAGE Publications Sage CA: Los Angeles, CA}
}

@inproceedings{zhang2024gaze,
  title={In gaze we trust: Comparing eye tracking, self-report, and physiological indicators of dynamic trust during hri},
  author={Zhang, Yinsu and Yadav, Aakash and Hopko, Sarah K and Mehta, Ranjana K},
  booktitle={Companion of the 2024 ACM/IEEE International Conference on Human-Robot Interaction},
  pages={1188--1193},
  year={2024}
}

@article{albert2010reliability,
  title={Reliability of self-reported awareness measures based on eye tracking},
  author={Albert, William and Tedesco, Donna},
  journal={Journal of Usability Studies},
  volume={5},
  number={2},
  pages={50--64},
  year={2010},
  publisher={Usability Professionals' Association Bloomingdale, IL}
}

@article{thilderkvist2024current,
  title={On current limitations of online eye-tracking to study the visual processing of source code},
  author={Thilderkvist, Eva and Dobslaw, Felix},
  journal={Information and Software Technology},
  volume={174},
  pages={107502},
  year={2024},
  publisher={Elsevier}
}

@article{nasteski2017overview,
  title={An overview of the supervised machine learning methods},
  author={Nasteski, Vladimir},
  journal={Horizons. b},
  volume={4},
  number={51-62},
  pages={56},
  year={2017}
}

@article{hoffmann2019benchmarking,
  title={Benchmarking in classification and regression},
  author={Hoffmann, Frank and Bertram, Torsten and Mikut, Ralf and Reischl, Markus and Nelles, Oliver},
  journal={Wiley Interdisciplinary Reviews: Data Mining and Knowledge Discovery},
  volume={9},
  number={5},
  pages={e1318},
  year={2019},
  publisher={Wiley Online Library}
}

@inproceedings{susmitha2024mitigating,
  title={Mitigating Driver Fatigue: A Multisensory Approach with Advanced Machine Learning Techniques},
  author={Susmitha, Arika Asha and Akshay, Gopu and Reddy, Marri Sahasra and Pagadala, Pavan Kumar and Kumar, Chanda Raj and Pinapatruni, Sree Lakshmi},
  booktitle={International Conference on Communications and Cyber Physical Engineering 2018},
  pages={937--945},
  year={2024},
  organization={Springer}
}

@article{li2025applications,
  title={Applications of Machine Learning in Assessing Cognitive Load of Uncrewed Aerial System Operators and in Enhancing Training: A Systematic Review},
  author={Li, Qianchu and Molloy, Oleksandra and El-Fiqi, Heba and Eves, Gary},
  journal={Drones},
  volume={9},
  number={11},
  pages={760},
  year={2025},
  publisher={MDPI}
}

@article{basnet2025real,
  title={Real-time cognitive workload assessment using non-intrusive methods: a systematic review},
  author={Basnet, Niosh and Zahabi, Maryam},
  journal={Ergonomics},
  pages={1--26},
  year={2025},
  publisher={Taylor \& Francis}
}

@article{bilbao2021objective,
  title={Objective and subjective evaluation of saccadic eye movements in healthy children and children with neurodevelopmental disorders: A pilot study},
  author={Bilbao, Carmen and Pi{\~n}ero, David P},
  journal={Vision},
  volume={5},
  number={2},
  pages={28},
  year={2021},
  publisher={MDPI}
}

@article{pandey2021temporal,
  title={Temporal and spatial feature based approaches in drowsiness detection using deep learning technique},
  author={Pandey, Nageshwar Nath and Muppalaneni, Naresh Babu},
  journal={Journal of Real-Time Image Processing},
  volume={18},
  number={6},
  pages={2287--2299},
  year={2021},
  publisher={Springer}
}

@article{zemblys2019gazenet,
  title={gazeNet: End-to-end eye-movement event detection with deep neural networks},
  author={Zemblys, Raimondas and Niehorster, Diederick C and Holmqvist, Kenneth},
  journal={Behavior research methods},
  volume={51},
  number={2},
  pages={840--864},
  year={2019},
  publisher={Springer}
}

@article{cheng2024appearance,
  title={Appearance-based gaze estimation with deep learning: A review and benchmark},
  author={Cheng, Yihua and Wang, Haofei and Bao, Yiwei and Lu, Feng},
  journal={IEEE Transactions on Pattern Analysis and Machine Intelligence},
  volume={46},
  number={12},
  pages={7509--7528},
  year={2024},
  publisher={IEEE}
}

@book{matloff2017statistical,
  title={Statistical regression and classification: from linear models to machine learning},
  author={Matloff, Norman},
  year={2017},
  publisher={Chapman and Hall/CRC}
}

@inproceedings{alnuaimi2024overview,
  title={An overview of machine learning classification techniques},
  author={Alnuaimi, Amer FAH and Albaldawi, Tasnim HK},
  booktitle={BIO Web of Conferences},
  volume={97},
  pages={00133},
  year={2024},
  organization={EDP Sciences}
}

@article{kadam2020regression,
  title={Regression techniques in machine learning \&applications: A review},
  author={Kadam, Vidya S and Kanhere, Shweta and Mahindrakar, Shrikant},
  journal={Int. J. Res. Appl. Sci. Eng. Technol},
  volume={8},
  number={10},
  pages={826--830},
  year={2020}
}

@article{skaramagkas2021review,
  title={Review of eye tracking metrics involved in emotional and cognitive processes},
  author={Skaramagkas, Vasileios and Giannakakis, Giorgos and Ktistakis, Emmanouil and Manousos, Dimitris and Karatzanis, Ioannis and Tachos, Nikolaos S and Tripoliti, Evanthia and Marias, Kostas and Fotiadis, Dimitrios I and Tsiknakis, Manolis},
  journal={IEEE reviews in biomedical engineering},
  volume={16},
  pages={260--277},
  year={2021},
  publisher={IEEE}
}

@article{kaczorowska2021interpretable,
  title={Interpretable machine learning models for three-way classification of cognitive workload levels for eye-tracking features},
  author={Kaczorowska, Monika and Plechawska-W{\'o}jcik, Ma{\l}gorzata and Tokovarov, Mikhail},
  journal={Brain sciences},
  volume={11},
  number={2},
  pages={210},
  year={2021},
  publisher={MDPI}
}

@inproceedings{skaramagkas2021cognitive,
  title={Cognitive workload level estimation based on eye tracking: A machine learning approach},
  author={Skaramagkas, Vasileios and Ktistakis, Emmanouil and Manousos, Dimitris and Tachos, Nikolaos S and Kazantzaki, Eleni and Tripoliti, Evanthia E and Fotiadis, Dimitrios I and Tsiknakis, Manolis},
  booktitle={2021 IEEE 21st International Conference on Bioinformatics and Bioengineering (BIBE)},
  pages={1--5},
  year={2021},
  organization={IEEE}
}

@inproceedings{rudmann2003eyetracking,
  title={Eyetracking in cognitive state detection for HCI},
  author={Rudmann, Darrell S and McConkie, George W and Zheng, Xianjun Sam},
  booktitle={Proceedings of the 5th international conference on Multimodal interfaces},
  pages={159--163},
  year={2003}
}

@article{saiz2021analysis,
  title={Analysis of the learning process through eye tracking technology and feature selection techniques},
  author={S{\'a}iz-Manzanares, Mar{\'\i}a Consuelo and P{\'e}rez, Ismael Ramos and Rodr{\'\i}guez, Adri{\'a}n Arnaiz and Arribas, Sandra Rodr{\'\i}guez and Almeida, Leandro and Martin, Caroline Fran{\c{c}}oise},
  journal={Applied Sciences},
  volume={11},
  number={13},
  pages={6157},
  year={2021},
  publisher={MDPI}
}

@article{ye2023supporting,
  title={Supporting traditional handicrafts teaching through eye movement technology},
  author={Ye, Li and Yang, Simin and Zhou, Xueyan and Lin, Yuxi},
  journal={International Journal of Technology and Design Education},
  volume={33},
  number={3},
  pages={981--1005},
  year={2023},
  publisher={Springer}
}






\end{document}